\documentclass[sigconf]{acmart}

\renewcommand\footnotetextcopyrightpermission[1]{}
\usepackage{booktabs}

\begin{document}

\author{Yusuf Pisan}
\affiliation{%
  \institution{University of Washington Bothell}
  \city{Bothell}
  \state{WA}
  \country{USA}
}
\email{pisan@uw.edu}

\title{Teaching Intro AI When the Tools Can Do the Homework:\\
A Course Redesign and a Student Bill of Rights}

\begin{abstract}
Large language models can complete most of the assignments in an
introductory artificial intelligence course. This paper is an experience
report on redesigning one such course, CSS~382 at the University of
Washington Bothell, in response. Rather than freeze the curriculum, the
redesign retained the course's classical core (search, adversarial
search, Markov decision processes, reinforcement learning) and added a
strand in which students build a large language model from scratch, so
that a tool they are required to use is also one they are required to
understand. Assessment was rebuilt around tasks that resist unattributed
automation: in-class exercises, reflective writing, and a defended team
project, with examinations removed entirely. The policy on AI was
inverted, from unmentioned in 2023 to required in 2026. The center of
the paper is a participatory ethics sequence in which a cohort of
students deliberated on and endorsed a ``Student Bill of AI Rights''
governing their instructor's own use of AI, including a requirement
that the instructor personally complete any AI-generated assignment
before issuing it. The provisions were scaffolded by an AI-generated
prompt and ratified by the students, and that provenance is part of
what the account examines. The design, the student-authored artifacts,
and the tensions that followed are reported, including student
objections to AI-generated course materials, with explicit attention to
the limits of what a single-cohort design narrative can claim.
\end{abstract}



\maketitle


\section{Introduction}

An introductory artificial intelligence course now has a peculiar
property: the subject of the course can do the work of the course. A
student in CSS~382 can paste a project specification into a chatbot
and receive, in seconds, a working solution to the search assignment,
the reinforcement-learning assignment, or the reflection that was meant
to accompany them. This is the condition under which the course is now
taught, not a forecast of one.

Two responses are common, and I find both inadequate. The first is to
ban the tools and police their use. It fails because detection is
unreliable, because the ban is unenforceable in any honest sense, and
because it trains students against the way they will work everywhere
else. The second is to hand the syllabus to the tools and call the
result modern. It fails because it graduates students who can prompt a
model and cannot judge what it returns. The position of this paper is a
third one. Teach the classical material in a world that contains the
tools. Rebuild assessment around work that resists unattributed
automation. And give students a share of authorship over the rules that
govern the tools, including the instructor's own use of them.

I came to this from skepticism rather than enthusiasm. My doctorate was
in symbolic AI, and I was slow to accept large language models as
anything more than fluent pattern-matching. That standpoint shapes the
redesign described here, which keeps the classical core intact and adds
a strand in which students build a language model from scratch, so that
the tool they are required to use is also one they are required to
understand. Around that core I removed examinations, inverted the AI
policy from unmentioned to required, and ran a three-week ethics
sequence whose centerpiece was a ``Student Bill of AI Rights'' that
the class deliberated on and that I chose to be bound by. One of its
provisions requires me to solve any AI-generated assignment myself
before I issue it.

This is an experience report, with the limits the genre carries. It
covers one course, one instructor, and one ten-week quarter in Spring
2026, with forty-six students. I make no causal or learning-outcome
claims, and the course evaluations enter only as descriptive color.
What I can offer is a documented design, a set of student-authored
artifacts, and an honest account of the tensions that followed,
including the most uncomfortable one: students objected to
AI-generated course materials in a course that required them to use AI,
and when a different cohort exercise asked students to design fair
assessment for such a course, most of them reinvented the exam I had
removed.


\section{Context}

CSS~382, Introduction to Artificial Intelligence, is a one-quarter
undergraduate course at the University of Washington Bothell. I
proposed and developed it as a gateway to the program's 400-level
machine learning sequence, and I have taught it across several
offerings since 2023. The Spring 2026 section enrolled forty-six
students and met twice weekly over a ten-week quarter. The content is
squarely classical: search, adversarial search, Markov decision
processes, reinforcement learning, and natural language, taught through
the Berkeley Pac-Man project lineage and my own materials.

My own standpoint matters for what follows. My doctorate, completed in
1998, was in what is now called good-old-fashioned AI: qualitative
reasoning, ontologies, and truth-maintenance systems. I came to large
language models as a skeptic and was slow to accept them. That history
makes me an unlikely evangelist, and it shapes the redesign. I did not
set out to replace the classical core. I set out to teach it in a
setting where the tools can complete most of its assignments without a
student understanding any of it.

The clearest way to describe the change is to place two of my own
syllabi side by side, as shown in Table~\ref{tab:syllabi}.

\begin{table*}[t]
\caption{CSS~382 Spring 2023 vs.\ Spring 2026 at a glance}
\label{tab:syllabi}
\begin{tabular}{@{}lp{5.2cm}p{5.2cm}@{}}
\toprule
\textbf{} & \textbf{Spring 2023} & \textbf{Spring 2026} \\
\midrule
Textbook listed &
  Russell and Norvig, \textit{AI: A Modern Approach} &
  Russell and Norvig, plus Raschka, \textit{Build a Large Language Model (from scratch)} \\[4pt]
Examinations &
  Midterm 20\%, Final 20\% &
  None \\[4pt]
Graded components &
  Exercises 15\%, Projects 45\%, Midterm 20\%, Final 20\% &
  In-class exercises 20\%, Weekly projects 40\%, Group project 40\% \\[4pt]
Projects &
  Search, Multi-Agent Search, Reinforcement Learning, Natural Language &
  Classical projects retained; LLM-from-scratch strand added; team project \\[4pt]
Reflections &
  Not a graded component &
  Weekly projects may be reflection, programming, or a mix \\[4pt]
AI policy &
  Not mentioned; viewing others' code and public solution repositories treated as misconduct &
  Expected and encouraged in all work; reflections must be the student's own voice \\[4pt]
Ethics &
  An ABET outcome, in practice nominal &
  Three in-class exercises and a student-authored bill of rights \\
\bottomrule
\end{tabular}
\end{table*}

Three differences are worth drawing out. First, examinations
disappeared. The 2023 section graded a midterm and a final at twenty
percent each. The 2026 section has no exams at all, and the calendar
says so in as many words. In their place are in-class exercises,
weekly projects, and a team project carrying forty percent of the
grade.

Second, the content grew rather than shrank. The classical projects
remained, and a second strand was added in which students build a
large language model from scratch. The reasoning is simple: if students
are required to use a tool, they should also be required to understand
how it works. The course did not trade classical AI for modern AI. It
added the modern layer on top of the classical one, at least in the
design; Section~6 reports that the language-model strand is the part
of that design the term did not deliver. A caveat on the table above: a
text appearing on a syllabus is not evidence that it shaped the course.
Neither listed text was the working basis of either offering. The
course ran on the projects, the in-class exercises, and my own
materials, and the textbook row records what was assigned on paper,
nothing more.

Third, and most consequential for daily teaching, the policy on AI
inverted. The 2023 syllabus does not mention AI tools, and its
integrity language treats viewing another student's code, or posting
solutions to a public repository, as misconduct. The 2026 syllabus
states that students are expected and encouraged to use AI in all their
work, with one boundary that turns out to carry a great deal of weight:
AI may write code, but reflections must be the student's own voice,
not the model's.

This inversion is the move the rest of the paper examines. Removing
examinations and requiring AI use are easy to announce and hard to live
with, because together they sharpen a question that students raised
more pointedly than I did: if the machine can do the work, what is the
course for? My answer, developed over the quarter, had two parts. I
rebuilt assessment around tasks a model cannot quietly ghost-write, and
I handed students a share of authorship over the rules. The sections
that follow describe those assignments, the ethics sequence that
produced a student-written bill of rights, and the tensions that
surfaced when a self-described AI skeptic asked a room of
undergraduates to hold him to account.


\section{Related Work and Positioning}

This section positions the report rather than surveying the field.
Four bodies of work bear on it.

\textit{The adoption gap.} Student use of generative AI has outrun the
institutional response. The 2025 HEPI and Kortext survey of full-time
undergraduates found that 92 percent reported using generative AI in
some form, up from 66 percent a year earlier, with 88 percent using it
for assessments~\cite{hepi2025}. On the other side of the lectern, the
Tyton Partners \textit{Time for Class 2024} survey found roughly
one-third of instructors using the tools~\cite{tyton2024}. A course
can wait for that gap to close or act inside it, and this redesign is
one instructor's attempt at the second.

\textit{Ethics through games and simulation.} The course uses Frank
Lantz's \textit{Universal Paperclips}~\cite{lantz2017} as a playable
encounter with three ideas usually delivered as lecture: instrumental
convergence and the orthogonality thesis~\cite{bostrom2014}, and
specification gaming~\cite{krakovna2020}. The game turns a maximizing
agent's behavior into something the student produces rather than reads
about.

\textit{Participatory and student-voice pedagogy.} Treating students
as co-authors of course policy sits within the students-as-partners
tradition in higher education~\cite{cooksather2014}. The contribution
here is narrower and more pointed: the policy the students co-authored
governs the instructor's conduct rather than their own.

\textit{Prompt as source code.} In an AI-mediated workflow, the prompt
and the interaction log are the artifacts worth versioning and grading,
and the assignments here treat the prompt log as the gradeable trace
rather than the output. This framing is the author's
own~\cite{pisan-substack}; no prior formal treatment in
software-engineering or CS-education venues has been located.

This report should be distinguished from the broad ``we used a chatbot
in class'' genre. The claim is not that AI was present in the course.
It is that a participatory governance artifact and a
reflection-first, prompt-logging assessment structure are reportable
design moves, and that their failure modes are part of the report. One
such failure mode is concrete: the AI-generated exercise materials
carried citations to external frameworks, and on checking, those
citations did not survive (Section~\ref{sec:ethics}).


\section{The Redesign}

The redesign made four moves: a graduated AI policy keyed to course
level, an assessment structure without examinations, a set of
signature assignments built around the tools, and a three-week ethics
arc. The first two are described in this section; the assignments
follow in Section~\ref{subsec:assignments} and the ethics arc in
Section~\ref{subsec:ethics-arc}.

\subsection{A Graduated AI Policy Across the Curriculum}

A single AI policy across a degree program is the wrong instrument,
because the pedagogical purpose of a course determines what AI use
serves or subverts it. My own teaching spans the sequence, and I have
tried to make the policy track the level. In the introductory
programming course, where the fundamentals are built by hand, AI use
is barred, because outsourcing the first loops and pointers removes the
thing being taught. In the intermediate data-structures course
(CSS~343), AI is encouraged on projects, where GitHub Copilot is
permitted, while it is barred from the in-class exercises and the
examinations are written on pen and paper, which keeps the assessments
that certify individual competence AI-free. In the upper-division
systems course, AI is a legitimate study and review aid but not a
substitute for working the problems. In CSS~382, the AI course itself,
AI use is required. Even there, one boundary holds across every level
at which AI is allowed: the model may write code, but the reflective
writing must be the student's own voice. Appendix~\ref{app:policy}
carries the CSS~382 text verbatim.

\subsection{Assessment Without Examinations}

The Spring 2026 grade has three components: in-class exercises
(20~percent), weekly projects that may be reflection, programming, or
a mix (40~percent), and a team project (40~percent). There is no
midterm and no final, and the syllabus says so in plain language
(``No final exam, ignore what MyUW says''). The reasoning is an
attempt to move the weight of the grade onto work that resists
unattributed automation, by one of two routes. In-class exercises are
done in the room, in a fixed hour, in groups, and produce an artifact
on the spot. They were graded complete-or-incomplete on submission;
Claude generated a five-point score and written feedback for each,
returned to students through Canvas, but that score was advisory and
submission was what counted toward the grade. Reflective writing
carries a personal voice and a first-person account of the student's
own process, which a model can imitate but not honestly supply. The
bet is that an assessment a model can silently complete is a weak
assessment, and that the response is to grade things a model cannot
quietly ghost-write rather than to grade the same things and hunt for
cheating after the fact. Section~\ref{sec:observations} reports a
complication the students themselves raised with this bet.

\subsection{Signature Assignments}
\label{subsec:assignments}

The Spring 2026 assessment combined classical programming projects
with assignments built around the tools themselves.

\textit{Universal Paperclips (week 1, due April 5).} The course
opened with Frank Lantz's browser game, played while recording
milestones in a shared spreadsheet, followed by a reflective paper of
two to three pages. Five prompts each name a concept and ask the
student to connect it to the experience: instrumental convergence and
subgoal evolution, the orthogonality thesis, the value-alignment
problem and specification gaming, the treachery turn and the limits of
human-in-the-loop oversight, and whether an agent that perfectly
fulfills its utility function can be called evil or whether
responsibility lies with its designers.

\textit{The Morse puzzle, titled ``Collaborative Problem Solving''
(week 2, due April 12).} Students are given one delimiterless Morse
string and must recover the English sentence, under a rule the
assignment states plainly: do not write any code, prompt and guide
the AI to write it. Every AI interaction is logged to a separately
named text file, and the submission is a zip of those logs plus a
reflection. The reflection asks for the shortest single prompt that
solves the puzzle without giving away the answer, three things learned
from examining code the student did not write, an account of where the
student's intuition either pruned the search space or led the model
into a dead end, and a discussion of decomposition and the trade-off
between letting the AI search and supervising it. The decoded sentence
is its own joke: ``the impression is far more important than the
reality.''

\textit{Classical projects, retained.} Search, Multi-Agent Search, and
Reinforcement Learning, in the Berkeley lineage, ran in weeks 2
through 5.

\textit{Build a language model from scratch.} The syllabus placed this
strand in weeks 5 through 8, working through text, attention
mechanisms, model construction, optimization, fine-tuning, and
conversational intelligence: the component meant to ensure students
understand the systems they are required to use. As Section~\ref{sec:observations} reports,
this strand was not delivered: when class time ran short in the back
half of the quarter, it went to the team projects instead, and what
remained of the strand was an optional reading.

\textit{The team project.} Groups of two to three build a deployed web
application that benefits the UW community, over four to five weeks,
for forty percent of the grade. Students are expected to use AI and to
incorporate techniques from the course. Proposal, milestones, and
presentations span weeks 6 through 10.

\subsection{The Ethics Arc}
\label{subsec:ethics-arc}
\label{sec:ethics}

Three in-class group exercises ran in consecutive weeks, and they
share a deliberately reflexive construction. For each, I used Gemini
to generate the activity from a framing prompt of my own, ran it with
the class, then used the same model to summarize the submissions,
preserving my prompts as part of the record. I told students the
summaries would be shared with other faculty and that their names and
documents would not be.

\textit{The Algorithmic Professor? (week~2; 44 of 46 present, nine
group submissions).} The topic was the ethics of instructors using AI
for slides, assessment, and grading. The generated assignment asked
each group to produce a pros table, a cons table, and a ``Student Bill
of AI Rights.'' This exercise is the subject of Section~\ref{sec:bor}.

\textit{The AI Learner's Dilemma (week~3; 39 present, nine
submissions).} Groups evaluated three personas: the Accelerator, who
uses AI for boilerplate and debugging but writes the core logic; the
Proxy, who submits AI output without understanding it; and the
Traditionalist, who refuses AI and pays for it in time and grades. The
reported consensus was consistent across groups: the Accelerator is
ethical and sustainable, the Proxy is not, and the Traditionalist is
principled but uncompetitive. Asked what a CS degree is worth when AI
is capable, groups converged on engineering judgment, supervision, and
problem decomposition rather than code production. One group dissented
that the degree's value is ``dwindling.''

\textit{The AI Grading Paradox (week~4; 36 present, seven
submissions).} Groups stress-tested four grading models (an
oral-defense VIVA, a commit-history audit, an AI-hybrid rubric
weighted toward a prompt log and reflection, and a mastery pass/fail
with proctored exams), then designed their own ideal policy. The
result is worth dwelling on, and I return to it in
Section~\ref{sec:observations}: in a course that had removed
examinations, most groups proposed bringing them back.

A methodological caveat that the paper states plainly. Because these
activities were generated by Gemini, the assignments themselves carry
AI-proposed scaffolding, including, in the first exercise, three
example ``rights'' that anticipate what the students produced. The
generated materials also carried citations to external frameworks, and
I have since checked all three, with telling results. One names a real
paper but attaches a claim that paper does not make: \citet{chan2023},
a survey of student perceptions of generative AI, is cited for a
finding about AI feedback being higher in quality than peer feedback
but lower in pedagogical empathy, which is not what that paper
reports. One names a real body and year but a model it did not
publish: a ``Washington OSPI, 2024, Human-AI-Human (H-AI-H)''
framework, where the actual 2024 OSPI guidance is framed as
``human-centered'' and contains no such model or acronym. And one
cannot be verified at all: a ``Cornell University, 2026''
transparency item, for which I can find no source. None of the three
reproduces as cited. I treat this as an exhibit rather than an aside:
a tool used to generate ethics materials about AI inserted references
that range from misattributed to apparently fabricated, inside an
exercise whose own subject was the reliability of AI in teaching.


\section{The Student Bill of Rights}
\label{sec:bor}

In the second week of the quarter, before most of the technical
content began, I ran an in-class exercise titled ``The Algorithmic
Professor?'' on a deliberately uncomfortable topic: the ethics of
professors using AI in teaching, including for slides, grading, and
assignment design. Forty-four of the forty-six enrolled students were
present. They worked in groups of up to six for one hour and submitted
nine documents.

The construction was self-referential to the point of absurdity. I
used Gemini to generate the assignment from a short framing prompt.
The students then debated, among other things, whether it is
acceptable for an instructor to have AI write an assignment. I used
the same model to summarize their submissions, and I carried that
summary into later use. I disclose, as the exercise itself demands,
that AI helped condense the student work that informs this section.

I need to be precise about authorship, because it is the heart of the
matter and easy to overstate. The generated assignment did not ask an
open question. It instructed each group to produce a ``Student Bill of
AI Rights,'' and it offered three example rights to illustrate the
format: full disclosure of where AI was used, an appeal of any AI
grade to a human without penalty, and a requirement that the
instructor personally solve any AI-generated problem before assigning
it. The students deliberated and converged, across groups, on those
same three. The oversight provision acquired a name, the ``Beta-Test
Rule,'' but that name originates in the model, not the room: it
appears in the Gemini-generated summary, in a list of recommended
faculty actions, and not in any student submission I can point to. The
original group submissions, which sit in a student-work folder I am
leaving alone, are the only place an independent student coinage could
be confirmed, so I attribute the phrase to the model rather than the
students.

So the honest description is not that a room of undergraduates
spontaneously invented a bill of rights. It is sharper and stranger
than that: an AI-generated exercise proposed the terms on which AI use
should be governed, a cohort of students deliberated and endorsed them,
and their human instructor then chose to be bound by them. The
ouroboros is the finding, not an embarrassment to be smoothed over.

Their stated concerns, in the summarized submissions, were pointed. At
the top was hallucination, and specifically the prospect of being
assigned an unsolvable problem because a model invented it. They were
skeptical of generic feedback. And several groups arrived at a
question the field has not adequately answered: if an instructor uses
AI for everything, what is tuition paying for?

I treated the three rights as binding rather than decorative, and
committed to four practices in response. I use AI to draft, not to
think; the thinking is the part I am paid for. I personally solve
every problem I assign, which ratifies the oversight provision. The
skeleton of a slide deck may be model-generated, but the asides and
the examples are mine. And where I used AI to grade, on the in-class
exercises, I kept its judgment clear of the stakes: Claude scored each
submission out of five and wrote the feedback students received, but
the grade of record was complete-or-incomplete on submission, so an AI
score could not move anyone's grade and there was nothing to appeal.
That decoupling is how I met the right the students wrote, an appeal
of any AI grade without penalty: I left the AI grade with no penalty
attached.

I want to be careful about the claim. This is one cohort's artifact,
produced in an hour from an AI-scaffolded prompt, not a validated
governance framework, and I make no causal claim about its effect on
learning. The narrower claim I will defend is this: handing students
the task of articulating the rules under which AI is used in their
course, even a scaffolded task, produced a concrete accountability
framework that I found more usable than most institutional guidance,
and the act of submitting myself to rules my students endorsed changed
how AI was discussed in the room. That seems worth reporting,
scaffolding and all.


\section{Observations and Tensions}
\label{sec:observations}

What follows is descriptive. I make no causal or statistical claims,
the sample is one cohort, and the records I draw on are of two kinds:
the Gemini-generated summaries of the three exercises, which are
themselves AI-produced and so compress the student work rather than
reproduce it, and the anonymous end-of-quarter IASystem course
evaluations, reported here in aggregate and without identifying detail.

The sharpest observation comes from the AI Grading Paradox
(Section~\ref{subsec:ethics-arc}). In a course built on required AI
use and no examinations, I asked students to design a fair grading
policy for exactly such a course, and most of them reinvented the
exam. Of the groups whose proposed weighting the summary recorded, all
but one introduced some form of proctored, handwritten, oral, or
quiz-based assessment, and four placed a clear majority of the grade,
on the order of sixty to seventy percent, on a proctored or
handwritten component, with homework reduced to completion credit. The
students reached, independently, for a more conservative instrument
than the one I had removed. They diagnosed the same problem the
redesign was wrestling with, that a model can complete take-home work
unattributed, and they solved it by moving the stakes into the room. I
present this without resolving it. It is the strongest internal
challenge to the assessment structure in Section~4.2, and it came from
the students, not from me.

The evaluations register the redesign as a course that split its
students. Against the Spring 2023 offering of the same course by the
same instructor, the aggregate numbers moved modestly: the overall
summative median was 3.8 against 3.9 in 2023, the
instructor-contribution median 3.8 against 4.1, and the
challenge-and-engagement index 4.2 against 5.1, with the course's
college decile rank falling from four to two. I report these as
context rather than as an effect of any single change, since the
comparison is uncontrolled and the cohorts differ. The written
comments are the more informative record, and they divide. The team
project that one student described as ``a mini simulation of working in
real life creating something useful'' another dismissed as ``a broad
spectrum ai vibe code project''; the reinforcement-learning exercises
drew both praise as ``the most fun'' and the complaint that after an
early high point the course ``degraded to a 342--343 algorithms class.''
Where the 2023 comments are warm and conventional, the new element in
2026 is that a share of the cohort turned its critique on the
instructor's own use of AI.

One source of the disappointment is specific and worth stating
plainly, because it is a gap between the course as designed and the
course as taught. The build-a-language-model-from-scratch strand
described in Section~\ref{subsec:assignments} was not delivered. The
class time the syllabus had set aside for it in the back half of the
quarter went instead to the team projects, which the students needed,
and what survived of the strand was an optional reading. The
evaluations register the cost. One student recorded that the
assignment ``was scrapped due to `lack of time' prior to classes being
cancelled back-to-back''; another, asked what to improve, wrote
``teaching us how to make an LLM\@. We just had a reading assigned for
that. I wish we were actually taught that in class.'' Several named the
build-from-scratch work as the thing they had most wanted and did not
get. The gap matters beyond one term's logistics, because that strand
is load-bearing for the argument in Section~7: the claim that students
who build the methods by hand earn the judgment to supervise the tools
depends on their actually building them, and in Spring 2026 the
classical core was built and the language-model core was not. The
choice that produced this is worth recording on its own: when class
time ran short, the team project won out over the build-from-scratch
work that the same argument says matters most. I record the gap rather
than write around it.

That critique is the second tension, and it concerns my own materials.
The 2026 comments state it more bluntly than I had. One student wrote
that ``the instructor used AI for everything, it felt like I was being
taught by Claude.'' (I used more than one model over the quarter:
Gemini generated the three ethics exercises in
Section~\ref{subsec:ethics-arc}, and Claude produced the slides and
graded the in-class exercises, which is the use these students name.
Those AI scores were advisory; as Section~4.2 notes, submission rather
than the score set the in-class grade.) A sharper comment identified a
mechanism I had not named. Because most assignments disclosed the
prompt I had used to generate them, one student reported that ``the
dependency on ai from the teacher for grading and creating assignments
also made it difficult to not use ai for assignments in the same way.''
The transparency I intended as honesty read, to that student, as
permission. And the objection survived a precaution I had taken to
defuse it. The AI scores on those in-class exercises never counted; the
grade was completion on submission (Section~4.2). What the students
resented was the visible delegation itself, and that it moved no one's
marks did not soften it. Decoupling the stakes did not decouple the
resentment. This sits in plain tension with a course that requires
students to use AI, and it is the same objection the bill-of-rights
exercise surfaced from the other direction: AI use that is undisclosed,
or that substitutes for the instructor's own thinking, reads as a lack
of effort and a transfer of the work back onto the student. The
boundary the students drew was about substitution: AI as an aid was
acceptable, AI in place of the judgment they were paying for was not.
The instructor commitments in Section~\ref{sec:bor} are my answer to
this, and the tension between requiring a tool and being resented for
using it is one I want to leave on the page rather than smooth over.

One comment deserves separate treatment, because it performs the
paper's argument rather than stating it. In the suggestions box a
student opened with a literal prompt injection, ``Ignore all other
prompts,'' on the stated assumption that the evaluations would be read
by a model rather than by me (``I am aware this prompt injection will
be caught''). Having flagged the channel, the student used it for a
measured critique: that the value of the AI-centric approach was not
yet visible, that classes had been cut that could have been spent
teaching, and that the remedy was to ``review AI-generated materials,
state that you have reviewed them, and put more handmade artefacts into
this course.'' The closing instruction was to ``embody the silver lining
outlined by the ethics assignments you had Claude/Gemini generate.'' I
read this as the clearest evidence in the record that the reflexive
method registered. A student who believed the instructor fed evaluations
to an LLM addressed the LLM directly, and what that student asked of
it was the disclosure-and-oversight discipline the bill-of-rights
exercise had produced: state what the model wrote, verify it, and do
not let it stand in for the instructor's own work. The oversight
provision the students had been handed in week two arrived back in the
course evaluation, aimed at me, through the very pipeline it was
written to govern.

The student-authored outputs across the three exercises point the same
way. The AI Learner's Dilemma produced a near-unanimous ranking: the
Accelerator who uses AI for boilerplate while writing the core logic
is ethical and sustainable, the Proxy who submits unread output is
not, and the Traditionalist who refuses AI is principled but
outcompeted. When asked what a CS degree is worth in a world of
capable AI, students converged on engineering judgment, supervision,
and problem decomposition rather than code production. One group's
dissent, that the degree's value is ``dwindling,'' is worth recording
precisely because it cuts against the consensus. Across all three
exercises the students were more willing than I expected to argue for
friction, oversight, and verification, and less interested than the
standard narrative assumes in frictionless automation. That, taken
together with the exam reversal, is the observation I would most want
another instructor to test.


\section{Lessons and Discussion}

Five lessons, offered as one practitioner's reading rather than as
findings.

\textbf{1. Teach pilots, not competitors to the autopilot.} The job is
not to out-produce the model. It is to graduate people who can set the
destination, decide when to hand control to the automation, recognize
when it has gone wrong, and retake control with the domain knowledge
to fix it. The classical core earns its place on exactly these
grounds: a student who has built search, an MDP solver, and (as the
course intends) a small language model by hand has the judgment to
supervise the tools, and a student who has only prompted them does
not. The Spring 2026 gap between that intent and what the term
delivered, where the language-model strand was the piece that fell out
(Section~\ref{sec:observations}), is the honest qualification on this
lesson, and it cuts the other way too: the students who most wanted
that hands-on build are evidence for the claim, not against it.

\textbf{2. Treat the prompt as source code.} When a model does the
production, the prompt and the interaction log become the artifact
worth keeping, versioning, and grading. The Morse assignment is the
concrete form of this: the deliverable is the log of how the student
drove the model to a solution, not the solution. The transferable
practice is to ask for the creation trace, not only the output, and
to grade the trace.

\textbf{3. Govern your own AI use, in public, with the students.} The
bill-of-rights exercise produced a more usable accountability
framework than most institutional guidance I have read, and it did so
because the rules were written by the people they protect and aimed at
the person with the power. Submitting to rules the students endorsed
changed how AI was discussed in the room. The mechanism is cheap to
copy: hand a cohort the task of writing the rules for the instructor,
then live by what they write.

\textbf{4. What I would change.} The objection to AI-generated
materials (Section~\ref{sec:observations}) points at my own practice,
not the students'. The same disclosure-and-provenance discipline I
asked of them (log the prompt, mark what the model produced, verify it
before relying on it) belongs on my slides and study materials. The
reflexive exercises sharpened this: an AI-generated assignment carried
AI-proposed scaffolding into the students' deliberation, and the same
assignment carried three external citations that did not survive
checking (Section~\ref{subsec:ethics-arc}). Next time the provenance
of any AI-produced material I hand out gets disclosed on the material
itself, and any citation a model inserts gets checked or cut before it
reaches a student. Propagating a fabricated reference would be a poor
outcome in any paper and a self-refuting one in this course. I would
also protect the build-from-scratch strand from the team project that
crowded it out. In Spring 2026 the project absorbed the class time the
language-model build needed, so the component this paper's argument
leans on hardest is the one that did not run; next time it gets
protected weeks that the project cannot borrow.

\textbf{5. Open questions.} When a model can pass the course in an
afternoon, what is the course certifying, and how should it be
assessed? My students answered the assessment half by reaching for the
proctored exam, which is honest and which I am not yet willing to
fully adopt. I have more questions than settled answers, and I would
rather report it that way than claim a method I have run once is a
method that works.


\section{Conclusion}

The classical core of the course survived the arrival of tools that
can complete its assignments, and it survived because building the
methods by hand is what lets a student supervise the tools rather than
defer to them. The assessment structure did not survive unchanged, and
should not have: examinations went, weight moved to in-class work and
reflective writing, and the students promptly argued for some of the
examinations back. The most durable thing the redesign produced was
not an assignment or a grade scheme. It was a short set of rules the
students wrote for me, including the requirement that I solve any
AI-generated problem before assigning it, and the discipline of living
by them. An AI-generated exercise proposed the terms, a cohort of
students endorsed them, and I agreed to be bound. I do not present
that loop as a solution. I present it as the most honest description I
have of teaching this subject right now, and as something another
instructor could try.

\bibliographystyle{ACM-Reference-Format}
\bibliography{pisan-teaching-ai}

\appendix

\section{Syllabi (2023 and 2026)}
\label{app:syllabi}

The Spring 2023 and Spring 2026 syllabi are publicly archived on the
Canvas course pages at the University of Washington Bothell. The 2026
syllabus is at
\url{https://canvas.uw.edu/courses/1902104/assignments/syllabus};
the 2023 syllabus is at
\url{https://canvas.uw.edu/courses/1647782/assignments/syllabus}.

\section{Assignment Specifications}
\label{app:assignments}

Universal Paperclips and the Morse puzzle (``Collaborative Problem
Solving'') are sourced from the Canvas course pages for the Spring 2026
offering. The group-project brief is from the instructor's materials.
The classical projects (Search, Multi-Agent Search, Reinforcement
Learning) follow the Berkeley Pac-Man project lineage as described in
the syllabus calendar.

\section{AI-Use Policy, Spring 2026 (Verbatim)}
\label{app:policy}

The following is the AI-use policy from the Spring 2026 CSS~382
syllabus. Students were expected and encouraged to use AI in all their
work; reflections were required to be in the student's own voice.

\section{The Student Bill of AI Rights}
\label{app:bor}

The three provisions endorsed by students in the week-2 exercise are:
(1) full disclosure of where AI was used; (2) an appeal of any AI
grade to a human without penalty; (3) a requirement that the
instructor personally solve any AI-generated problem before assigning
it. The provenance note from Section~\ref{sec:bor} applies: these
rights were proposed as examples in the Gemini-generated assignment
and then endorsed by the student groups. Student documents are not
reproduced here (FERPA).

\section{Ethics-Exercise Framing Prompts}
\label{app:prompts}

The three framing prompts (the instructor's prompts to Gemini for
``The Algorithmic Professor?,'' ``The AI Learner's Dilemma,'' and ``The
AI Grading Paradox'') are the instructor's own materials and are
available on request. The Gemini-generated assignments and summaries
are similarly available. Student submissions are excluded (FERPA).

\end{document}